\RequirePackage[hyphens]{url}

\documentclass[%
reprint,
superscriptaddress,
 hidelinks,
 amsmath,amssymb,
 aip,
floatfix,
]{revtex4-2}

\usepackage{graphicx}
\usepackage{dcolumn}
\usepackage{bm,booktabs}
\usepackage{amsthm}
\usepackage{multirow}
\usepackage{color, soul}
\usepackage{hyperref}
\usepackage{tikz}

\tikzstyle{io} = [rectangle, minimum width=3cm, minimum height=1cm, text centered, text width=3cm, draw=black]
\tikzstyle{arrow} = [thick,->,>=stealth]

\newcommand{\AffilSLAC}{\affiliation{%
 SLAC National Laboratory,
 Menlo Park, California 94025, USA
}}

\newcommand{\AffilWisc}{\affiliation{%
 University of Wisconsin-Madison, 
 Madison, Wisconsin 53706, USA
}}

\newcommand{\elmo}{\mbox{ELM-O} }
\newcommand{\elmop}{\mbox{ELM-O's} }

\begin{document}

\preprint{APS/123-QED}

\title{Automatic Identification of Edge Localized Modes in the DIII-D Tokamak}

\author{Finn H. O'Shea}%
 \email{foshea@slac.stanford.edu}
 \AffilSLAC
\author{Semin Joung}
    \AffilWisc
\author{David R. Smith}
 \AffilWisc
\author{Ryan Coffee}
 \AffilSLAC

\date{\today}

\begin{abstract}
Fusion power production in tokamaks uses discharge configurations that risk producing strong Type I Edge Localized Modes.  
The largest of these modes will likely increase impurities in the plasma and potentially damage plasma facing components such as the protective heat and waste divertor.  
Machine learning-based prediction and control may provide for online mitigation of these damaging modes before they grow too large to suppress.
To that end, large labeled datasets are required for supervised training of machine learning models.
We present an algorithm that achieves 97.7\% precision when automatically labeling Edge Localized Modes in the large DIII-D tokamak discharge database.
The algorithm has no user controlled parameters and is largely robust to tokamak and plasma configuration changes.
This automatically-labeled database of events can subsequently feed future training of machine learning models aimed at autonomous Edge Localized Mode control and suppression.
\end{abstract}

\maketitle


\section{\label{sec:intro}Introduction}

High confinement (H-mode) operation modes are the mode of choice for fusion power production in future tokamak fusion systems \cite{clark2006nuclear}.  These modes are targeted for the "highest performance" operation at ITER \cite{smith_path_2010}.   
Unfortunately, these operational modes also produce so-called Edge Localized Modes (ELMs) at the plasma boundary that result in damaging heat loads at the divertor and other plasma facing components \cite{federici_key_2003, counsell_boundary_2002}.
These modes also increase the impurity of the plasma through contamination by erosion of those components \cite{clark2006nuclear}.  
Management of ELMs has been named a ``critical issue for ITER'' \cite{wade_physics_2009}, where fusion power production requires plasma density levels that produce Type I ELMs \cite{clark2006nuclear}.  
These ELMs are expected to occur up to once per second at ITER \cite{federici_key_2003}.

On the other hand, the ELMs in a so-called ``ELMy'' H-mode also provide a mechanism for density and impurity control of the core plasma \cite{evans_suppression_2004, zohm_physics_1996}.  
For this reason, it is desirable to control the ELMs during H-mode operation rather than strictly eliminate them.  
There is an existing literature on control of ELMs that primarily focuses on adapting plasma parameters and system design at time scales covering many ELMs to reduce the harmful effects of the worst of them \cite{evans_suppression_2004, zohm_edge_1996, zohm_physics_1996, wade_physics_2009, snyder_elms_2004, saibene_characterization_2005, loarte_characteristics_2003, rapp_elm_2002, rhee_mechanism_2012, degeling_magnetic_2003, lang_elm_2003, liang_magnetic_2013, maingi_edge-localized-mode_2009, shousha_design_2022}.

There is mounting optimism in an alternative to plasma parameter selection or design; real-time predictive plasma control systems could preemptively identify that an ELM will occur in the future $<$100 ms and autonomously prescribe corrective actions to ameliorate the damaging consequences.  
Rich with sensor systems that have well defined semantic meaning, tokamaks are good candidates for monitoring and active control by the data-driven models of machine learning. 
For example, a neural network was used to predict disruption of the ASDEX Upgrade tokamak ``far away from the desired operational space (H-mode, high-$\beta$)'' \cite{pautasso_prediction_2001}. 
There is a mounting literature on machine learning-based plasma control \cite{snipes_mhd_2011, kates-harbeck_predicting_2019, churchill_deep_2020, boyer_toward_2021, esquembri_real-time_2018, fu_machine_2020, abbate_data-driven_2021, zhu_scenario_2021, jalalvand_real-time_2022}.

A critical difficulty with training supervised machine learning models in the context tokamak reactors is the lack of labeled ELM events.  
Herein, we address this difficulty by describing an algorithm for autonomously identifying ELM events in a large data set of shots from the DIII-D tokamak \cite{luxon_design_2002}.  
The performance of this algorithm is evaluated against a set of ELMs that were hand-labeled by human experts and demonstrates near perfect precision and recall.  
Deployed on shots that were not selected by experts for ELM-potential, where recall is much more difficult to evaluate, the algorithm demonstrated a precision of 97.7\%.

To our knowledge there is no known algorithm or system for predicting an oncoming Type I ELM. 
However, some precursors to these ELMs have been observed, such as: an increase in broadband magnetic and density turbulent fluctuations \cite{zohm_edge_1996}, precursor oscillations in the electron edge temperature \cite{zohm_edge_1996}, presence of enhanced MHD activity between ELMs \cite{saibene_characterization_2005}, and wave-like edge intensity fluctuations and various magnetic signals \cite{sechrest_two-dimensional_2012}.  
There are also known magnetic precursors for Type III ELMs \cite{zohm_physics_1996}.


There are a variety of methods used to control ELMs at many-ELM timescale such as: resonant magnetic perturbations \cite{evans_suppression_2004}, increasing the impurity in the plasma \cite{rapp_elm_2002}, molecular beam injection \cite{rhee_mechanism_2012}, vertical motion of the plasma \cite{degeling_magnetic_2003}, pellet injection \cite{lang_elm_2003}, or magnetic topology changes via microwave power input \cite{liang_magnetic_2013}.  
Alternatively, the character and expected harm of the ELMs can be addressed via careful parameter selection \cite{saibene_characterization_2005} or modification of the vacuum vessel materials \cite{maingi_edge-localized-mode_2009}.

From the above survey of work in the field of ELM physics, it appears plausible that there is sufficient information in the tokamak diagnostic signals to predict the appearance of an individual ELM.  
The potential benefit of such predictions is quite large given the statement, ``disruption avoidance and onset warning $\ldots$ would have a very favourable implication on the use of a full W clad divertor in ITER.'' \cite{federici_key_2003}

Recent work uses an ELM detector based exclusively on a single filterscope signal \cite{shousha_design_2022}, further described elsewhere \cite{eldon_controlling_2017}.  Those authors state that ``the ELM detector [just referenced] should be replaced by a more robust ELM detector that does not rely on calibration-specific tuning'' \cite{shousha_design_2022}.  Here, we use additional signals, but with no user-controlled parameters, to create just such an ELM detector that has high precision for more than 5 years worth of shots recorded at DIII-D.  
This system can be used to create a training data set for machine learning models, to detect, classify, and ultimately predict ELMs in real-time based on time series streaming data from tokamak diagnostic sensors.

Other recent, complementary, work has trained a U-net for ELM discovery at the KSTAR tokamak \cite{song_development_2022}.  Therein, the authors train the neural network on a single filterscope signal using 10 shots for training (and validation) and 2 shots for testing and compare the performance of their model to several others.  This model does not contain any user-controlled parameters.  We call this related work complementary because it is an example of how labeled data can be used to create high quality models for plasma science.  We will compare the reported performance of this U-net with the performance of our model after describing the latter.


\section{\label{sec:methods}ELM Observer (ELM-O): An Algorithm for Labeling ELMs in Time Series Data}

\begin{figure}[ht]
    \centering
    \begin{tikzpicture}[node distance=2cm]
        \node[] (data_in) {DIII-D archive};
        
        \node[below of=data_in] (fil) {3 filterscopes};
        \node[left of=fil, below of=fil] (int) {2 interferometers};
        \node[right of=fil] (bes) {BES};
        
        \node[io, below of=int] (up_samp) {up-sample \\ each signal \\ to 1 MHz} ;
        \node[right of=up_samp, below of=int] (blank0) {};
        \node[io, right of=blank0] (rescale) {rescale half of \\ channels by 2} ;

        \node[io, below of=up_samp] (diff) {first order \\ time derivative} ;
        \node[right of=diff] (blank1) {};
        \node[io, right of=blank1] (average) {average \\ 64 channels} ;

        \node[io, below of=diff] (perc) {percentile filter \\ by signal \\ $>$0.997};
        \node[right of=perc] (blank2) {};
        \node[io, right of=blank2, below of=blank1] (thresh) {threshold \\ $>$1};

        \node[io, below of=blank2] (smooth) {convolution \\ smoothing, \\ 100 $\mu s$ kernel} ;

        \node[below of=smooth] (blank3) {};
        \node[io, left of=blank3] (int_elm) {ELM candidate \\ if either \\ interferometer \\ signals ELM};
        \node[io, right of=blank3] (bes_elm) {ELM candidate \\ if BES threshold \\ exceeded};

        \node[below of=int_elm] (blank4) {};
        \node[io, right of=blank4] (fil_elm) {ELM candidate \\ if at least 2 of 3 \\ filterscopes \\ signal ELM};

        \node[io, below of=fil_elm] (is_elm) {ELM \\ if candidate \\ in all three \\ signal types};
    
        \draw [arrow] (data_in) -| (int);
        \draw [arrow] (data_in) -- (fil);
        \draw [arrow] (data_in) -| (bes);
        
        \draw [arrow] (int) -- (up_samp);
        \draw [arrow] (fil) |- (up_samp);
        \draw [arrow] (bes) -- (rescale);

        \draw [arrow] (up_samp) -- (diff);
        \draw [arrow] (rescale) -- (average);
        
        \draw [arrow] (diff) -- (perc);
        \draw [arrow] (average) -- (thresh);

        \draw [arrow] (perc) -- (smooth);
        \draw [arrow] (thresh) -- (smooth);

        \draw [arrow] (smooth) -| (int_elm);
        \draw [arrow] (smooth) -| (bes_elm);
        \draw [arrow] (smooth) -- (fil_elm);

        \draw [arrow] (int_elm) |- (is_elm);
        \draw [arrow] (fil_elm) -- (is_elm);
        \draw [arrow] (bes_elm) |- (is_elm);

    \end{tikzpicture}
    \caption{Data flow through the \elmo algorithm.}
    \label{fig:data_flow_diagram}
\end{figure}
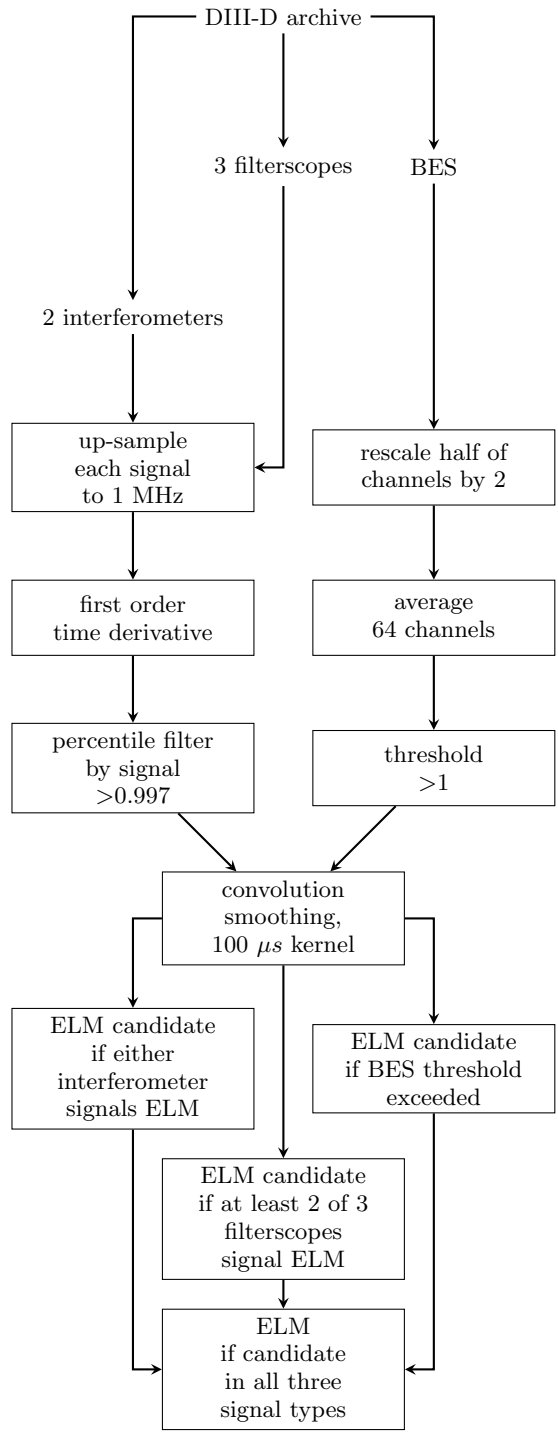

To locate the ELMs in time, we use diagnostic signals from the tokamak: two interferometric measurements of the plasma density (interferometers) \cite{van_zeeland_phase_2004}, three measurements of atomic emission lines (filterscopes) \cite{brooks_filterscopes_2008}, and 64 channels from the Beam Emission Spectroscopy system (BES) \cite{mckee_beam_1999}.
These three systems are sensitive to ELMs of all types, not just Type I ELMs.
All three instrument types have different sampling rates as shown in Table \ref{tab:sr}.  
We consider time periods within a shot when all the instruments are reporting data, we call this the concurrent sampling window.  
The data is processed by the \elmo algorithm to produce labeled ELMs.  
The algorithm is described diagrammatically in Fig. \ref{fig:data_flow_diagram}.

\begin{table}[htbp]
\caption{\label{tab:sr}%
Sampling rates for the instruments used in this work. 
}
\begin{tabular}{lc}
\toprule
instrument & sample rate (kS/s)  \\\midrule
filterscope    & 50      \\
interferometer     & 100      \\
BES    & 1000      \\
\bottomrule
\end{tabular}
\end{table}

The factor of 20 difference in sampling rate between the filterscopes and the BES mean we have to ``align'' the data during the concurrent sampling window.  
To do this, we perform a discrete cosine transform (DCT) \cite{ahmed:1974} on the interferometer and filterscope data, append zeros to the end of the transformed vector, and inverse transform (IDCT) back to the time-domain.
This operation produces ``spectrum preserving'' interpolation to the same time-domain sample points as the high-sampling BES signal.
We note this approach could also be used to decimate BES data as well, e.g. one could effectively choose a middle point in sampling frequency between the disparate sources.

\begin{figure}[t]
\centering
\includegraphics[width=\linewidth]{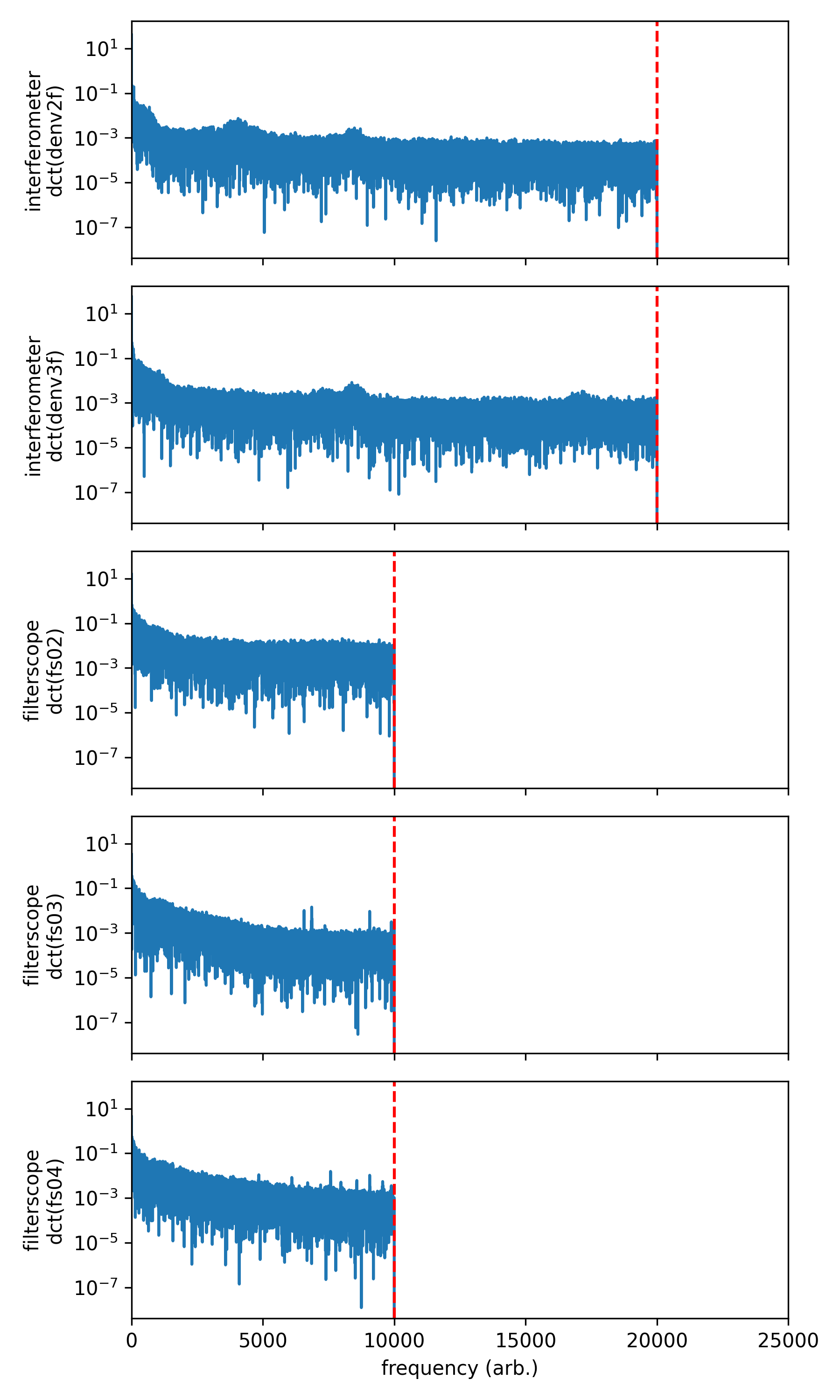}
\caption{\label{fig:dct_space} Spectrum of a 200 ms window of the interferometer and filterscope signals.  The dashed vertical line shows the highest frequency component of the DCT that exists, all higher frequency components are set identically to zero.}
\end{figure}

\begin{figure}[t]
\centering
\includegraphics[width=\linewidth]{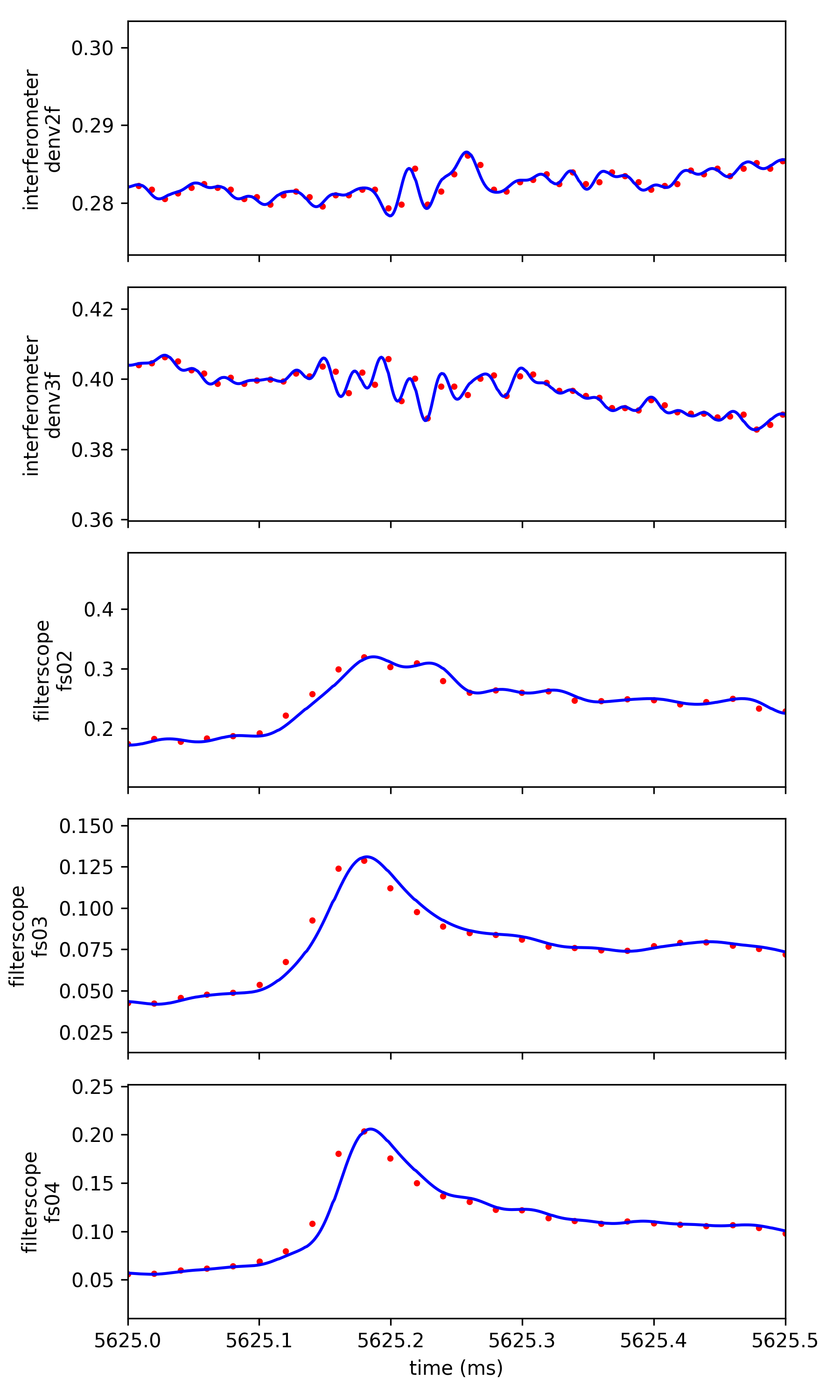}
\caption{\label{fig:before_and_after_dct} 0.5 ms time series from a tokamak shot showing an ELM.  The top two rows are interferometer signals sampled at 100 kS/s, the bottom three rows are filterscope signals sampled at 50 kS/s.  The red circles show the measurements before ``up-sampling'' to 1 MS/s, the blue lines show the signals after ``up-sampling.''}
\end{figure}

Typically, we perform this ``up-sampling'' routine on time series that are 200 ms long, or 200k points in the BES signal.  
The spectra of some example signals to be up-sampled are shown in Fig. \ref{fig:dct_space}.  
An example of the the signals before and after this transformation are shown in Fig. \ref{fig:before_and_after_dct}.  
Despite the sharp cut-off where the spectrum goes to zero, the ``spectrum preserving'' up-sampled signals are quite faithful to the originals.  
The largest change appears to be that the up-sampled filterscope signals are delayed by approximately 10~$\mu$s.

The BES data is reduced to a single time series.  
First, the latter 32 channels have half the range of the former 32 channels and so the signals from those channels are correspondingly scaled by 2.  
Second, all 64 channels are averaged together.  
The above procedure creates six channels for detecting ELMs when this single BES channel is combined with the 2 interferometer channels and 3 filterscope channels.

The principle behind the detection algorithm is that, because each of the different diagnostics detects different reactions of the plasma to ELMs, they should all be treated separately and then combined at the end.
In the BES signal, candidates are labeled via simple threshold.  
If the signal is above threshold $t$ (out of approximately 10 V), the point is labeled an ELM candidate.
The interferometer and filterscope signals are differentiated numerically using a first-order time difference.  
From each differentiated signal, the 1 - $\eta$ largest points are labeled as candidates, where the same $\eta$ is used for all 5 of the signals to define the threshold percentile.

All six signals are now binary time series in which a one indicates an ELM candidate in that signal.  
The signals are individually convolved with a window function 100~$\mu$s long to prevent a ``near miss'' between signals.  
This also helps produce fewer, more uniform regions (in time) of candidates instead of a greater number of smaller regions.
A candidate is declared in the interferometer signals if either signal contains a candidate.  
A candidate is declared in the filterscope signals if any two of the three signals contains a candidate.  
A candidate is declared in the BES signal directly, based on the average of the channels as described above.
Finally, any point that has a candidate in all three diagnostics is declared an ELM.

\begin{figure}[htbp]
\centering
\includegraphics[width=\linewidth]{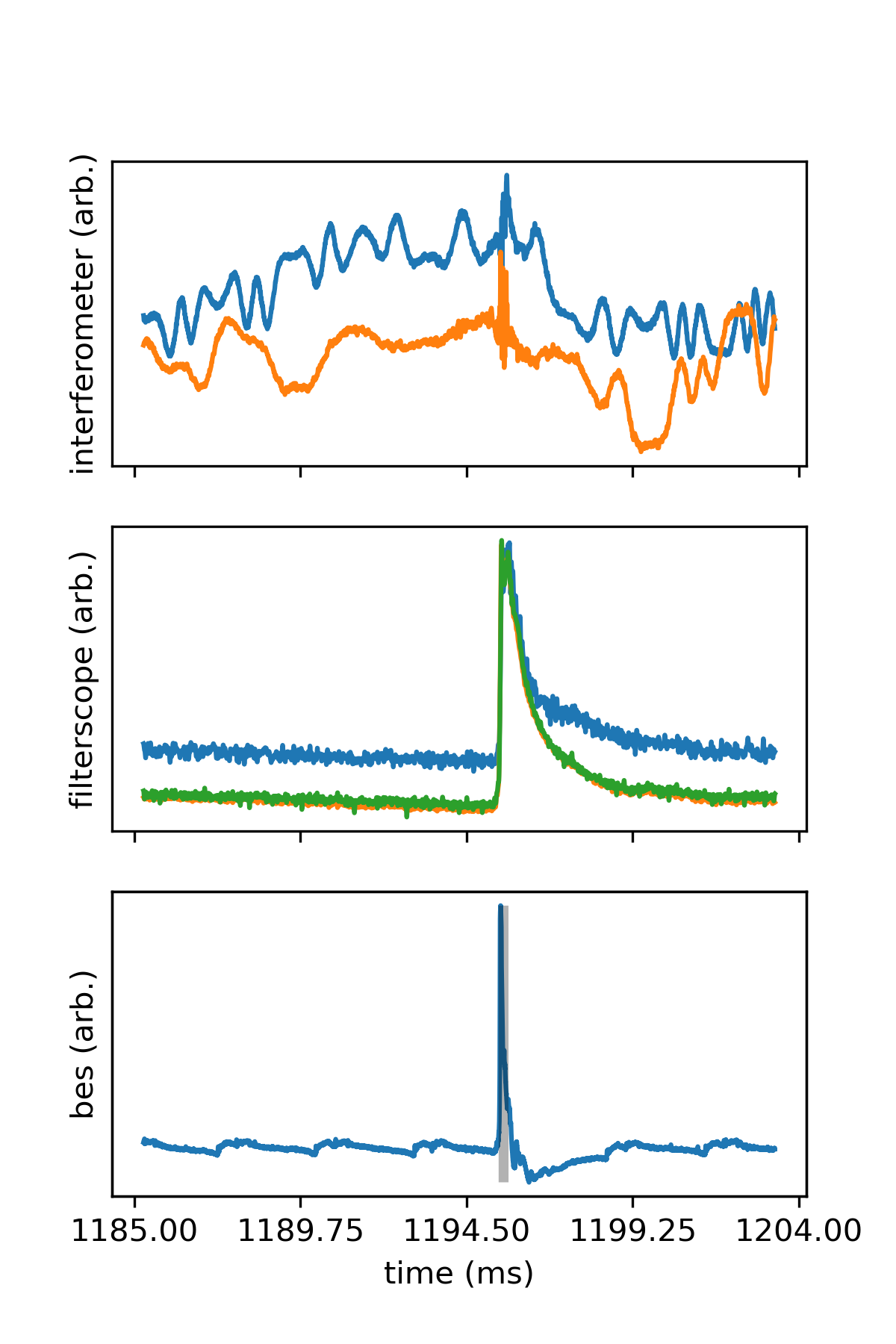}
\caption{\label{fig:elm_example} Example of one of the hand labeled ELMs.  
The top row shows the interferometer signals, the middle row the filterscope signals, and the bottom row the mean of the BES signals.  
The shaded region in the bottom row is the region hand-labeled as an ELM.}
\end{figure}

\begin{figure}[tbp]
\centering
\includegraphics[width=\linewidth]{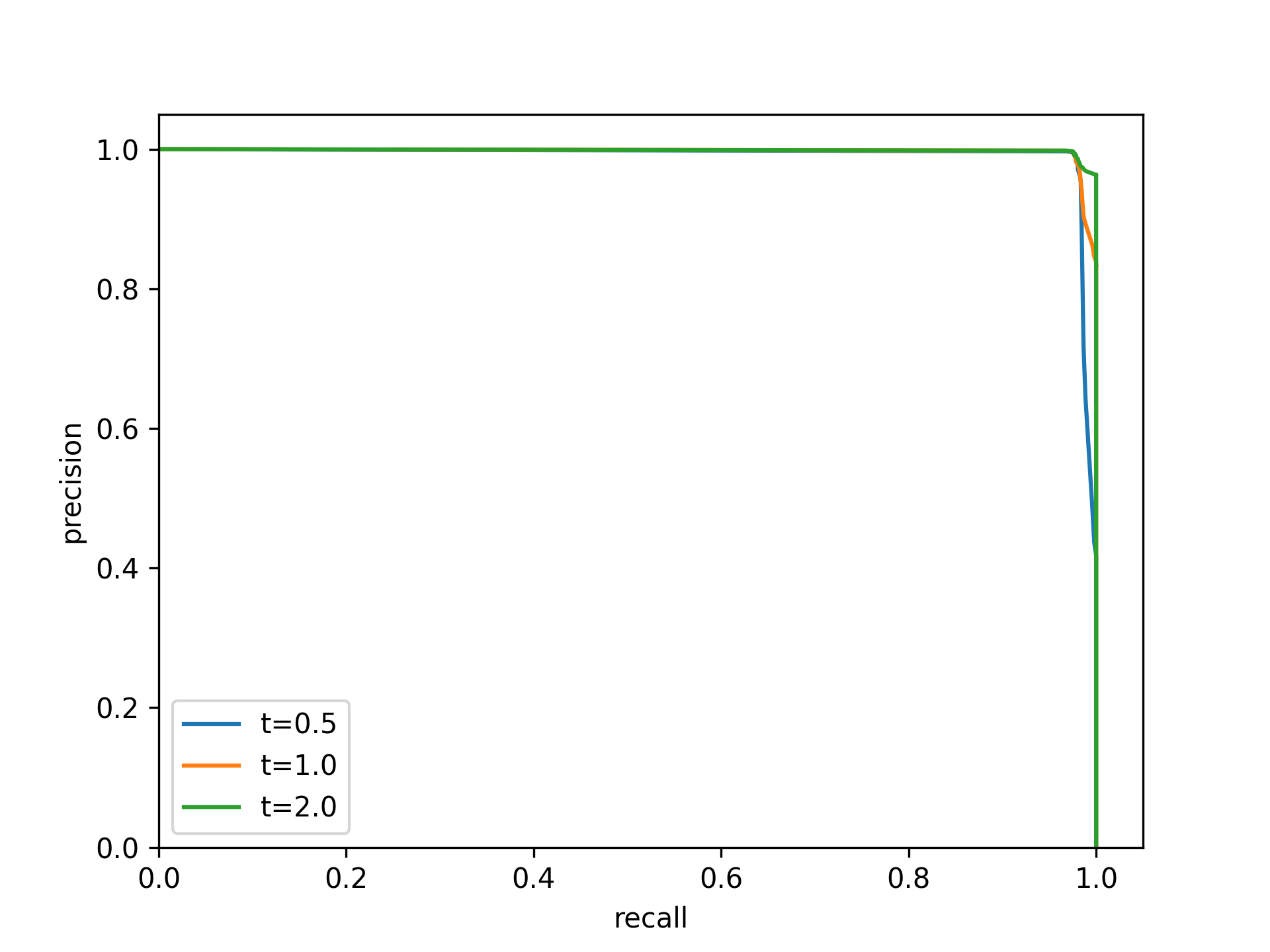}
\caption{\label{fig:pr_curve} Precision-recall curve for the \elmo algorithm.  The maximum AUC when $t=1$~V is 0.971 and occurs for a percentile-threshold of 0.997.}
\end{figure}

To tune the values of $t$ and $\eta$, we use a database of 972 ELMs that were hand labeled. 
This dataset consists of manually trimmed regions that should contain only one ELM and a smaller region labeled to be the ELM.  
An example of the data used to detect ELMs is shown in Fig. \ref{fig:elm_example}.  
The trimmed regions are of varying length from a few milliseconds to tens of milliseconds.  
In order to rank the thresholds, $t$ and $\eta$, we use the Area Under the Curve (AUC) metric of the precision-recall curve \cite{James:2013}.  
A plot of the precision-recall curve as a function of $\eta$ and $t$ is shown in Fig. \ref{fig:pr_curve}.

We define true positives, false positives, and false negatives in the following way.  
For each distinct time-span the algorithm labels as an ELM, a true positive is scored if there is any overlap with the hand-labeled ELM.  
If there is no overlap with the hand-labeled region, this is scored as a false negative.  
A false positive is scored when the algorithm labels a region to contain an ELM that does not contain a hand-labeled ELM.  
Note that it is possible for the algorithm to both miss the true ELM (false negative) and label a non-ELM as an ELM (false positive) in a single example.  
In fact, it is possible to have many false positives in a single example.  
In practice this is only a problem with very permissive labeling of ELMs.

\setlength{\tabcolsep}{5mm} 
\begin{table}[htbp]
\caption{\label{tab:pr}%
Algorithm Performance on the hand labeled data set.  Only the highest AUC $\eta$ is shown for each threshold.
}
\begin{tabular}{ccccc}
\toprule
$t$ & $\eta$ & P & R & AUC  \\ \midrule
0.5 & 0.998 & 0.996 & 0.974 & 0.970   \\
1.0 & 0.997 & 0.995 & 0.976 & 0.971   \\
2.0 & 0.997 & 0.997 & 0.975 & 0.972   \\
5.0 & 0.200 & 0.998 & 0.961 & 0.959   \\
\bottomrule
\end{tabular}
\end{table}

Table \ref{tab:pr} shows the highest performing settings of $t$ and $\eta$.
When $t$ is 2 or below, it appears that there is little cause to be selective in it's value.
We selected $t=1$~V for convenience.

With a percentile-threshold of $\eta = 0.997$ ($t=1$~V), the algorithm returns 5 false positives and 23 false negatives and an AUC of 0.971.  
Inspection of these errors reveals that the 5 false positives are all cases where the human-labeler did not label a second ELM in the time window, i.e. they are true positives.  
Of the false negatives, 18 come from two shots where the timing system failed to appropriately line up the signals from all three diagnostic types.  
Thus, the precision of the algorithm is perfect, while the recall depends on the functioning of the signal digitizer time registration.  
Assuming improved inter-diagnostic timing registration, we expect that the algorithm will produce nearly ideal detection of ELMs similar to those labeled by hand.

We emphasize here that the percentile setting is not a user-controlled parameter.
It has been set via tuning the algorithm's performance on our test set and any changes made to this parameter will require re-evaluation of the algorithm performance on a different hand-labeled set.
As we shall see shortly, these settings (and the overall algorithm design) show similar performance on seven years worth of data at DIII-D.
Hand tuning of these values on a per-shot basis is not feasible as our goal is to discover, en mass, ELMs in a large data set of tens of thousands of shots at DIII-D.

\begin{figure}[htbp]
\centering
\includegraphics[width=\linewidth]{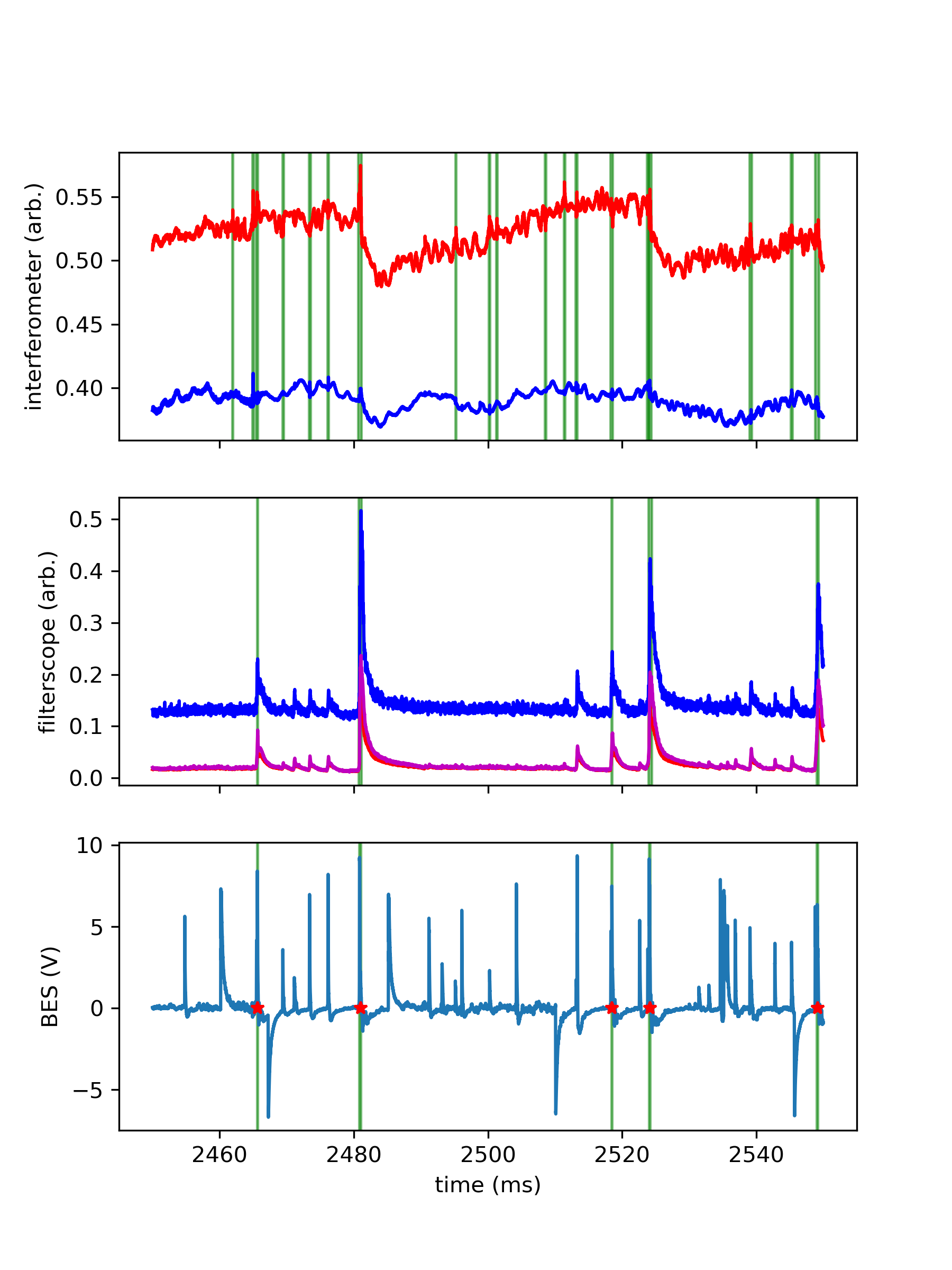}
\caption{\label{fig:missed_elms} 100 ms window from a period with both large and small ELMs.  The top row shows the interferometer signals, the middle row shows the filterscope signals, and the bottom row shows the mean BES signal.  The vertical regions in green show the parts of each signal labeled as ELMs, whereas the red stars in the bottom panel show the ELMs found by the \elmo algorithm.}
\end{figure}

Hereafter, we focus on precision as our evaluation metric for two reasons.  
Firstly, when creating a database of ELM events, our primary concern is making sure that the database contains only ELMs. 
Second, it is far easier for an expert to review potential ELMs that have already been labeled than it is to comb through the raw data looking for ELMs.
Potential ELM discovery is in fact the process we wish to automate with this work.

To further evaluate the performance of the algorithm we ran \elmo on randomly sampled time windows from a number of shots.
The shots were divided into two groups: shots (but not ELM events) that were previously used to produce the 972 hand labeled ELMs on which \elmop parameters were tuned (Group 1) and shots disjoint from these shots (Group 2).  
The group 1 data set contains 49 shots numbered within the range from 166433 (April 4, 2016) to 173224 (October 12, 2017).  
These shots had been selected prior to the work described here because the human expert expected to see type-I ELMs in these time series.  
For each shot, \elmo was shown a random 200~ms window and labeled all the ELMs in that window.  
This random window was selected from the entire time series recorded by the control system excluding 200~ms at the beginning and the end of the shot.  
We did not otherwise perform any selection for plasma status. 
This procedure produced 299 ELMs.  
An expert review of these ELMs found only 4 of them to be false positives.  
These ELMs came from two shots and are pairs labeling the same event described by an expert as "possibly a disruption of the plasma" separated by 200 to 300 $\mu s$.  
Nevertheless, the precision of \elmo is 98.7\% in this data set.

Group 2 ELMs are composed as follows. 
First, we identified a corpus of 9941 shots by filtering the entire DIII-D database with the following criteria: 1) the operational mode of the tokamak is set to ``plasma,'' 2) the discharge lasts for more than 1 second, and 3) the diagnostics we use in the algorithm show data in the database. 
Two hundred shots were randomly selected from this set numbered from 156562 (April 7, 2014) to 187328 (June 24, 2021) while excluding the entire range of shots included in the original 972 hand labeled ELMs.  
This resulted in 76 shots, 58 before the group 1 shots and 18 after.  
A 200~ms window is randomly selected from each of the shots in same way as for group 1.  
\elmo found 392 ELM candidates in this data.  
From this data, an expert found that 97.7\% (383/392) are indeed ELMs.

Inspection of the results on signals that contain smaller ELMs suggests that a different method be used to detect all the ELMs in a data series.  
In Figure \ref{fig:missed_elms} the filterscope shows some lower intensity ELMs between the larger ELMs.  
They also appear to show up in the BES signal.  
However, these smaller ELMs are not large enough to exceed the percentile-threshold, so they are not labeled as ELMs.  
Identifying these smaller ELMs automatically is the subject of on-going work.

Finally, we make a brief comparison between \elmo and the U-net model trained on data from KSTAR \cite{song_development_2022}.  Table \ref{tab:unet} shows the values for precision and recall for both models (reported in the other work as positive prediction rate and true positive rate, respectively).  Because we lack estimates for recall on our test data (as described earlier), we limit out comments to precision and simply note that \elmo has a smaller reduction in precision when deployed on unseen data.  We do not make much of this difference because the U-net was trained to identify smaller ELM's that we know \elmo does not identify.  Once again, we highlight the complementary nature of the two ELM detection methods: \elmo uses multiple signals to identify ELMs whereas the U-net uses a single filterscope signal and identifies ELMs "not [by] the intensity but the shape of the [ELM] peaks" \cite{song_development_2022}.

\begin{table}[htbp]
\caption{\label{tab:unet}%
Comparison between the performance of \elmo and the reported performance of the U-net described elsewhere \cite{song_development_2022}.  The values for the test set (shots 18396 and 29487) were interpolated from Figs. 9a and 10a, respectively.
}
\begin{tabular}{cccc}
\toprule
model & data & P & R  \\\midrule
\elmo & training & 0.995 & 0.976 \\
& group 1 & 0.987 & \-- \\
& group 2 & 0.977 & \-- \\\midrule
U-net & training & 0.924 & 0.935 \\
& 18396 & 0.84 & 0.96 \\
& 29487 & 0.87 & 0.88 \\
\bottomrule
\end{tabular}
\end{table}

\section{\label{sec:conclusions}Conclusions and Future Work}

``Big Data' has made machine learning much easier'' \cite{Goodfellow-et-al-2016} and we expect that `machine learning' will do the same for plasma science, but only if the inevitable hunger for labels can be fed.  
In this manuscript, we have shown that the algorithm \elmo is an effective way to automatically label ELMs in shots at DIII-D with no user-selected parameters.  
Testing the algorithm on shots spanning more than half a decade and with little regard for plasma parameter settings indicated that the false positive rate is less than 3\%.  
Lower false positive rate, less than 2\%, was achieved when a more careful selection of shots was made.  
We therefore conclude that \elmo is a good candidate for generating massive data sets of autonomously labeled ELMs, numbering in perhaps the millions.  
This is a dramatic improvement over current data sets that are painstakingly generated by human eye and number only in the thousands.

Since \elmo relies on percentile thresholds, it is robust to scale changes in the diagnostics, i.e. the gain setting of the filterscopes.  
However, this also means that \elmo will not label smaller ELMs in the presence of larger ELMs. 
See Fig. \ref{fig:missed_elms} for an example.  
Using \elmo-derived data, we are currently working on unsupervised machine learning methods for detecting ELMs of all shapes and sizes.
In addition, we are working on training a U-net model with the large labeled data sets we can generate using ELM-O.

\section{\label{sec:acks}Acknowledgements}

The authors would to thank R. Shousha and E. Kolemen for useful discussions.  
This work was supported by the Department of Energy, Office of Fusion Energy Science under Field Work Proposal 100636 ``Machine Learning for Real-time Fusion Plasma Behavior Prediction and Manipulation'' and Department of Energy Grant No. DE-SC0021157.  This material is based upon work supported by the U.S. Department of Energy, Office of Science, Office of Fusion Energy Sciences, using the DIII-D National Fusion Facility, a DOE Office of Science user facility, under Award(s) DE-FC02-04ER54698.

Disclaimer: This report was prepared as an account of work sponsored by an agency of the United States Government. Neither the United States Government nor any agency thereof, nor any of their employees, makes any warranty, express or implied, or assumes any legal liability or responsibility for the accuracy, completeness, or usefulness of any information, apparatus, product, or process disclosed, or represents that its use would not infringe privately owned rights. Reference herein to any specific commercial product, process, or service by trade name, trademark, manufacturer, or otherwise does not necessarily constitute or imply its endorsement, recommendation, or favoring by the United States Government or any agency thereof. The views and opinions of authors expressed herein do not necessarily state or reflect those of the United States Government or any agency thereof. 

\bibliography{main}

\end{document}